\input harvmac
\input epsf
\input amssym
\baselineskip 13pt




\def\cH{{\cal H}}
\def\cF{{\cal F}}
\def\dq{\delta q}
\def\xq{x_q}
\def\fin{f_{\infty}}
\def\N{{\cal N}}
\def\la{\langle}
\def\ra{\rangle}

\def\cH{{\cal H}}

\def\p{\partial}

\def\half{{1\over 2}}
\def\rar{\rightarrow}

\def\a{\alpha}
\def\b{\beta}

\def\l{\lambda}
\def\t{\tau}
\def\eps{\epsilon}

\def\zb{\overline{z}}

\def\vs{\vskip .1 in}



\newcount\figno
\figno=0
\def\fig#1#2#3{
\par\begingroup\parindent=0pt\leftskip=1cm\rightskip=1cm\parindent=0pt
\baselineskip=11pt
\global\advance\figno by 1
\midinsert
\epsfxsize=#3
\centerline{\epsfbox{#2}}
\vskip -21pt
{\bf Fig.\ \the\figno: } #1\par
\endinsert\endgroup\par
}
\def\figlabel#1{\xdef#1{\the\figno}}
\def\encadremath#1{\vbox{\hrule\hbox{\vrule\kern8pt\vbox{\kern8pt
\hbox{$\displaystyle #1$}\kern8pt}
\kern8pt\vrule}\hrule}}

\def\p{\partial}

\def\zb{\overline{z}}

\def\eps{\epsilon}

\def\zb{\overline{z}}

\def\cH{{\cal H}}

\def\IH{\Bbb{H}}
\def\IR{\Bbb{R}}

\baselineskip 14 pt

\lref\GalanteWTA{
  D.~A.~Galante and R.~C.~Myers,
  ``Holographic R\'enyi Entropies at Finite Coupling,''
[arXiv:1305.7191 [hep-th]].
}

\lref\KlebanovUF{
  I.~R.~Klebanov, S.~S.~Pufu, S.~Sachdev and B.~R.~Safdi,
  ``Renyi Entropies for Free Field Theories,''
JHEP {\bf 1204}, 074 (2012).
[arXiv:1111.6290 [hep-th]].
}

\lref\HungNU{
  L.~-Y.~Hung, R.~C.~Myers, M.~Smolkin and A.~Yale,
  ``Holographic Calculations of Renyi Entropy,''
JHEP {\bf 1112}, 047 (2011).
[arXiv:1110.1084 [hep-th]].
}

\lref\KrausVZ{
  P.~Kraus and F.~Larsen,
  ``Microscopic black hole entropy in theories with higher derivatives,''
JHEP {\bf 0509}, 034 (2005).
[hep-th/0506176].
}
\lref\HolzheyWE{
  C.~Holzhey, F.~Larsen and F.~Wilczek,
  ``Geometric and renormalized entropy in conformal field theory,''
Nucl.\ Phys.\ B {\bf 424}, 443 (1994).
[hep-th/9403108].
}

\lref\NishiokaHAA{
  T.~Nishioka and I.~Yaakov,
  ``Supersymmetric Renyi Entropy,''
[arXiv:1306.2958 [hep-th]].
}

\lref\ClossetRU{
  C.~Closset, T.~T.~Dumitrescu, G.~Festuccia and Z.~Komargodski,
  ``Supersymmetric Field Theories on Three-Manifolds,''
JHEP {\bf 1305}, 017 (2013).
[arXiv:1212.3388 [hep-th]].
}

\lref\BarnesBM{
  E.~Barnes, E.~Gorbatov, K.~A.~Intriligator, M.~Sudano and J.~Wright,
  ``The Exact superconformal R-symmetry minimizes tau(RR),''
Nucl.\ Phys.\ B {\bf 730}, 210 (2005).
[hep-th/0507137].
}

\lref\MyersJV{
  R.~C.~Myers, M.~F.~Paulos and A.~Sinha,
  ``Holographic studies of quasi-topological gravity,''
JHEP {\bf 1008}, 035 (2010).
[arXiv:1004.2055 [hep-th]].
}

\lref\AharonyNH{
  O.~Aharony, G.~Gur-Ari and R.~Yacoby,
  ``Correlation Functions of Large N Chern-Simons-Matter Theories and Bosonization in Three Dimensions,''
JHEP {\bf 1212}, 028 (2012).
[arXiv:1207.4593 [hep-th]].
}

\lref\MaldacenaSF{
  J.~Maldacena and A.~Zhiboedov,
  ``Constraining conformal field theories with a slightly broken higher spin symmetry,''
Class.\ Quant.\ Grav.\  {\bf 30}, 104003 (2013).
[arXiv:1204.3882 [hep-th]].
}

\lref\KlebanovTD{
  I.~R.~Klebanov, S.~S.~Pufu, S.~Sachdev and B.~R.~Safdi,
  ``Entanglement Entropy of 3-d Conformal Gauge Theories with Many Flavors,''
JHEP {\bf 1205}, 036 (2012).
[arXiv:1112.5342 [hep-th]].
}

\lref\ZhiboedovOPA{
  A.~Zhiboedov,
  ``On Conformal Field Theories With Extremal a/c Values,''
[arXiv:1304.6075 [hep-th]].
}

\lref\HofmanAR{
  D.~M.~Hofman and J.~Maldacena,
  ``Conformal collider physics: Energy and charge correlations,''
JHEP {\bf 0805}, 012 (2008).
[arXiv:0803.1467 [hep-th]].
}

\lref\CasiniKV{
  H.~Casini, M.~Huerta and R.~C.~Myers,
  ``Towards a derivation of holographic entanglement entropy,''
JHEP {\bf 1105}, 036 (2011).
[arXiv:1102.0440 [hep-th]].
}

\lref\OsbornCR{
  H.~Osborn and A.~C.~Petkou,
  ``Implications of conformal invariance in field theories for general dimensions,''
Annals Phys.\  {\bf 231}, 311 (1994).
[hep-th/9307010].
}

\lref\PetkouAD{
  A.~Petkou,
  ``Conserved currents, consistency relations and operator product expansions in the conformally invariant O(N) vector model,''
Annals Phys.\  {\bf 249}, 180 (1996).
[hep-th/9410093].
}

\lref\CasiniKT{
  H.~Casini and M.~Huerta,
  ``Entanglement entropy for the n-sphere,''
Phys.\ Lett.\ B {\bf 694}, 167 (2010).
[arXiv:1007.1813 [hep-th]].
}

\lref\PetkouVU{
  A.~C.~Petkou,
  ``C(T) and C(J) up to next-to-leading order in 1/N in the conformally invariant O(N) vector model for $2 < d < 4$,''
Phys.\ Lett.\ B {\bf 359}, 101 (1995).
[hep-th/9506116].
}

\lref\MyersTJ{
  R.~C.~Myers and A.~Sinha,
  ``Holographic c-theorems in arbitrary dimensions,''
JHEP {\bf 1101}, 125 (2011).
[arXiv:1011.5819 [hep-th]].
}

\lref\SolodukhinPK{
  S.~N.~Solodukhin,
  ``Entanglement entropy of round spheres,''
Phys.\ Lett.\ B {\bf 693}, 605 (2010).
[arXiv:1008.4314 [hep-th]].
}

\lref\DowkerRP{
  J.~S.~Dowker,
  ``Sphere Renyi entropies,''
J.\ Phys.\ A {\bf 46}, 225401 (2013).
[arXiv:1212.2098 [hep-th]].
}

\lref\DowkerBU{
  J.~S.~Dowker,
  ``Entanglement entropy for even spheres,''
[arXiv:1009.3854 [hep-th]].
}

\lref\DowkerYJ{
  J.~S.~Dowker,
  ``Entanglement entropy for odd spheres,''
[arXiv:1012.1548 [hep-th]].
}

\lref\KlebanovGS{
  I.~R.~Klebanov, S.~S.~Pufu and B.~R.~Safdi,
  ``F-Theorem without Supersymmetry,''
JHEP {\bf 1110}, 038 (2011).
[arXiv:1105.4598 [hep-th]].
}

\lref\BuchelSK{
  A.~Buchel, J.~Escobedo, R.~C.~Myers, M.~F.~Paulos, A.~Sinha and M.~Smolkin,
  ``Holographic GB gravity in arbitrary dimensions,''
JHEP {\bf 1003}, 111 (2010).
[arXiv:0911.4257 [hep-th]].
}

\lref\SolodukhinDH{
  S.~N.~Solodukhin,
  ``Entanglement entropy, conformal invariance and extrinsic geometry,''
Phys.\ Lett.\ B {\bf 665}, 305 (2008).
[arXiv:0802.3117 [hep-th]].
}

\lref\BelinDVA{
  A.~Belin, A.~Maloney and S.~Matsuura,
  ``Holographic Phases of Renyi Entropies,''
[arXiv:1306.2640 [hep-th]].
}

\lref\GubserZH{
  S.~S.~Gubser and I.~Mitra,
  ``Double trace operators and one loop vacuum energy in AdS / CFT,''
Phys.\ Rev.\ D {\bf 67}, 064018 (2003).
[hep-th/0210093].
}

\lref\EmparanGF{
  R.~Emparan,
  ``AdS / CFT duals of topological black holes and the entropy of zero energy states,''
JHEP {\bf 9906}, 036 (1999).
[hep-th/9906040].
}

\lref\MyersXS{
  R.~C.~Myers and A.~Sinha,
  ``Seeing a c-theorem with holography,''
Phys.\ Rev.\ D {\bf 82}, 046006 (2010).
[arXiv:1006.1263 [hep-th]].
}

\lref\VasilievBA{
  M.~A.~Vasiliev,
  ``Higher spin gauge theories: Star product and AdS space,''
In *Shifman, M.A. (ed.): The many faces of the superworld* 533-610.
[hep-th/9910096].
}

\lref\KlebanovJA{
  I.~R.~Klebanov and A.~M.~Polyakov,
  ``AdS dual of the critical O(N) vector model,''
Phys.\ Lett.\ B {\bf 550}, 213 (2002).
[hep-th/0210114].
}

\lref\GiombiMS{
  S.~Giombi and X.~Yin,
  ``The Higher Spin/Vector Model Duality,''
J.\ Phys.\ A {\bf 46}, 214003 (2013).
[arXiv:1208.4036 [hep-th]].
}

\lref\GiombiKC{
  S.~Giombi, S.~Minwalla, S.~Prakash, S.~P.~Trivedi, S.~R.~Wadia and X.~Yin,
  ``Chern-Simons Theory with Vector Fermion Matter,''
Eur.\ Phys.\ J.\ C {\bf 72}, 2112 (2012).
[arXiv:1110.4386 [hep-th]].
}

\lref\ChangKT{
  C.~-M.~Chang, S.~Minwalla, T.~Sharma and X.~Yin,
  ``ABJ Triality: from Higher Spin Fields to Strings,''
J.\ Phys.\ A {\bf 46}, 214009 (2013).
[arXiv:1207.4485 [hep-th]].
}

\lref\GiombiYVA{
  S.~Giombi, I.~R.~Klebanov, S.~S.~Pufu, B.~R.~Safdi and G.~Tarnopolsky,
  ``AdS Description of Induced Higher-Spin Gauge Theory,''
[arXiv:1306.5242 [hep-th]].
}

\lref\AharonyJZ{
  O.~Aharony, G.~Gur-Ari and R.~Yacoby,
  ``d=3 Bosonic Vector Models Coupled to Chern-Simons Gauge Theories,''
JHEP {\bf 1203}, 037 (2012).
[arXiv:1110.4382 [hep-th]].
}

\lref\MyersRU{
  R.~C.~Myers and B.~Robinson,
  ``Black Holes in Quasi-topological Gravity,''
JHEP {\bf 1008}, 067 (2010).
[arXiv:1003.5357 [gr-qc]].
}

\lref\RattazziGJ{
  R.~Rattazzi, S.~Rychkov and A.~Vichi,
  ``Central Charge Bounds in 4D Conformal Field Theory,''
Phys.\ Rev.\ D {\bf 83}, 046011 (2011).
[arXiv:1009.2725 [hep-th]].
}

\lref\PolandWG{
  D.~Poland and D.~Simmons-Duffin,
  ``Bounds on 4D Conformal and Superconformal Field Theories,''
JHEP {\bf 1105}, 017 (2011).
[arXiv:1009.2087 [hep-th]].
}

\lref\PolandEY{
  D.~Poland, D.~Simmons-Duffin and A.~Vichi,
  ``Carving Out the Space of 4D CFTs,''
JHEP {\bf 1205}, 110 (2012).
[arXiv:1109.5176 [hep-th]].
}

\lref\VichiUX{
  A.~Vichi,
  ``Improved bounds for CFT's with global symmetries,''
JHEP {\bf 1201}, 162 (2012).
[arXiv:1106.4037 [hep-th]].
}

\lref\ElShowkHT{
  S.~El-Showk, M.~F.~Paulos, D.~Poland, S.~Rychkov, D.~Simmons-Duffin and A.~Vichi,
  ``Solving the 3D Ising Model with the Conformal Bootstrap,''
Phys.\ Rev.\ D {\bf 86}, 025022 (2012).
[arXiv:1203.6064 [hep-th]].
}

\lref\KosTGA{
  F.~Kos, D.~Poland and D.~Simmons-Duffin,
  ``Bootstrapping the O(N) Vector Models,''
[arXiv:1307.6856 [hep-th]].
}

\lref\StanevQRA{
  Y.~S.~Stanev,
  ``Constraining conformal field theory with higher spin symmetry in four dimensions,''
[arXiv:1307.5209 [hep-th]].
}

\lref\AlbaYDA{
  V.~Alba and K.~Diab,
  ``Constraining conformal field theories with a higher spin symmetry in d=4,''
[arXiv:1307.8092 [hep-th]].
}

\lref\CasiniEI{
  H.~Casini and M.~Huerta,
  ``On the RG running of the entanglement entropy of a circle,''
Phys.\ Rev.\ D {\bf 85}, 125016 (2012).
[arXiv:1202.5650 [hep-th]].
}

\lref\DidenkoTV{
  V.~E.~Didenko and E.~D.~Skvortsov,
  ``Exact higher-spin symmetry in CFT: all correlators in unbroken Vasiliev theory,''
[arXiv:1210.7963 [hep-th]].
}

\lref\JafferisZI{
  D.~L.~Jafferis, I.~R.~Klebanov, S.~S.~Pufu and B.~R.~Safdi,
  ``Towards the F-Theorem: N=2 Field Theories on the Three-Sphere,''
JHEP {\bf 1106}, 102 (2011).
[arXiv:1103.1181 [hep-th]].
}

\lref\EmparanPM{
  R.~Emparan, C.~V.~Johnson and R.~C.~Myers,
  ``Surface terms as counterterms in the AdS / CFT correspondence,''
Phys.\ Rev.\ D {\bf 60}, 104001 (1999).
[hep-th/9903238].
}

\lref\BrownSJ{
  L.~S.~Brown and J.~P.~Cassidy,
  ``Stress Tensors and their Trace Anomalies in Conformally Flat Space-Times,''
Phys.\ Rev.\ D {\bf 16}, 1712 (1977)..
}

\lref\HerzogED{
  C.~P.~Herzog and K.~-W.~Huang,
  ``Stress Tensors from Trace Anomalies in Conformal Field Theories,''
[arXiv:1301.5002 [hep-th]].
}

\lref\LovelockYV{
  D.~Lovelock,
  ``The Einstein tensor and its generalizations,''
J.\ Math.\ Phys.\  {\bf 12}, 498 (1971)..
}

\lref\FursaevMP{
  D.~V.~Fursaev,
  ``Entanglement Renyi Entropies in Conformal Field Theories and Holography,''
JHEP {\bf 1205}, 080 (2012).
[arXiv:1201.1702 [hep-th]].
}

\lref\AnselmiAM{
  D.~Anselmi, D.~Z.~Freedman, M.~T.~Grisaru and A.~A.~Johansen,
  ``Nonperturbative formulas for central functions of supersymmetric gauge theories,''
Nucl.\ Phys.\ B {\bf 526}, 543 (1998).
[hep-th/9708042].
}

\lref\GreenTV{
  M.~B.~Green and M.~Gutperle,
  ``Effects of D instantons,''
Nucl.\ Phys.\ B {\bf 498}, 195 (1997).
[hep-th/9701093].
}

\lref\ErdmengerYC{
  J.~Erdmenger and H.~Osborn,
  ``Conserved currents and the energy momentum tensor in conformally invariant theories for general dimensions,''
Nucl.\ Phys.\ B {\bf 483}, 431 (1997).
[hep-th/9605009].
}

\lref\OsbornQU{
  H.~Osborn,
  ``N=1 superconformal symmetry in four-dimensional quantum field theory,''
Annals Phys.\  {\bf 272}, 243 (1999).
[hep-th/9808041].
}

\lref\AharonyNS{
  O.~Aharony, S.~Giombi, G.~Gur-Ari, J.~Maldacena and R.~Yacoby,
  ``The Thermal Free Energy in Large N Chern-Simons-Matter Theories,''
JHEP {\bf 1303}, 121 (2013).
[arXiv:1211.4843 [hep-th]].
}

\lref\SachdevPR{
  S.~Sachdev,
  ``Polylogarithm identities in a conformal field theory in three-dimensions,''
Phys.\ Lett.\ B {\bf 309}, 285 (1993).
[hep-th/9305131].
}

\lref\zyc{
K.~{\.Z}yczkowski, ``R{\'e}nyi extrapolation of shannon entropy,'' Open
  Systems \& Information Dynamics {\bf 10} (2003), no.~03 297--310.}

\lref\CalabreseQY{
  P.~Calabrese and J.~Cardy,
  ``Entanglement entropy and conformal field theory,''
J.\ Phys.\ A {\bf 42}, 504005 (2009).
[arXiv:0905.4013 [cond-mat.stat-mech]].
}

\lref\HeadrickZT{
  M.~Headrick,
  ``Entanglement Renyi entropies in holographic theories,''
Phys.\ Rev.\ D {\bf 82}, 126010 (2010).
[arXiv:1006.0047 [hep-th]].
}

\lref\RyuBV{
  S.~Ryu and T.~Takayanagi,
  ``Holographic derivation of entanglement entropy from AdS/CFT,''
Phys.\ Rev.\ Lett.\  {\bf 96}, 181602 (2006).
[hep-th/0603001].
}

\lref\LewkowyczNQA{
  A.~Lewkowycz and J.~Maldacena,
  ``Generalized gravitational entropy,''
[arXiv:1304.4926 [hep-th]].
}

\lref\NishiokaUN{
  T.~Nishioka, S.~Ryu and T.~Takayanagi,
  ``Holographic Entanglement Entropy: An Overview,''
J.\ Phys.\ A {\bf 42}, 504008 (2009).
[arXiv:0905.0932 [hep-th]].
}

\lref\CappelliVW{
  A.~Cappelli and A.~Coste,
  ``On The Stress Tensor Of Conformal Field Theories In Higher Dimensions,''
Nucl.\ Phys.\ B {\bf 314}, 707 (1989)..
}

\lref\CandelasGF{
  P.~Candelas and J.~S.~Dowker,
  ``Field Theories On Conformally Related Space-times: Some Global Considerations,''
Phys.\ Rev.\ D {\bf 19}, 2902 (1979)..
}

\lref\TakayanagiKG{
  T.~Takayanagi,
  ``Entanglement Entropy from a Holographic Viewpoint,''
Class.\ Quant.\ Grav.\  {\bf 29}, 153001 (2012).
[arXiv:1204.2450 [gr-qc]].
}

\lref\SwingleHGA{
  B.~Swingle,
  ``Structure of entanglement in regulated Lorentz invariant field theories,''
[arXiv:1304.6402 [cond-mat.stat-mech]].
}


\Title{\vbox{\baselineskip14pt
}} {\vbox{\centerline {A universal feature of CFT R\'enyi entropy}}}
\centerline{Eric Perlmutter\foot{E.Perlmutter@damtp.cam.ac.uk}}
\bigskip
\centerline{\it{DAMTP, Centre for Mathematical Sciences,  University of Cambridge}}
\centerline{${}$\it{Cambridge, CB3 0WA, UK}}

\baselineskip14pt

\vskip .3in

\centerline{\bf Abstract}
\vskip.2cm
We show that for a $d$-dimensional CFT in flat space, the R\'enyi entropy $S_q$ across a spherical entangling surface has the following property: in an expansion around $q=1$, the first correction to the entanglement entropy is proportional to $C_T$, the coefficient of the stress tensor vacuum two-point function, with a fixed $d$-dependent coefficient. This is equivalent to a similar statement about the free energy of CFTs living on $S^1\times \IH^{d-1}$ with inverse temperature $\b=2\pi q$. In addition to furnishing a direct argument applicable to all CFTs, we exhibit this result using a handful of gravity and field theory computations. Knowledge of $C_T$ thus doubles as knowledge of R\'enyi entropies in the neighborhood of $q=1$, which we use to establish new results in 3$d$ vector models at large $N$.

\Date{}

\newsec{Introduction}
The R\'enyi entropy is a measure of the degree of entanglement between two components of a tensor product Hilbert space. Upon forming a reduced density matrix $\rho$ by tracing over one subspace, one defines the R\'enyi entropy $S_q$ as
\eqn\dtg{S_q = {1\over 1-q}\log\Tr[\rho^q]}
where $q\neq 1$ is a positive, real parameter. In the limit $q\rar 1$, this reduces to the von Neumann entropy, otherwise known as the entanglement entropy, $S_{EE}\equiv \lim_{q\rar 1}S_q$. These quantities are UV divergent, but hiding behind the infinities are finite parts, free of ambiguities, that characterize the theory under study. These entropies have also proven quite valuable in an AdS/CFT context, as tests of the correspondence and as novel observables associated to regions of spacetime \refs{\RyuBV, \LewkowyczNQA}.
Much more can be said about this rapidly developing field, and we refer the reader to \refs{\CalabreseQY, \NishiokaUN, \HeadrickZT, \TakayanagiKG} and references therein for proper introductions to some of these ideas.

The aim of this work is to show that the R\'enyi entropy of generic conformal field theories in flat space enjoys a pleasantly simple and general relation to the coefficient of the stress tensor vacuum two-point function, $C_T$. 

Let us give some context. We henceforth consider a $d$-dimensional CFT living in flat space and sitting in its ground state, and we take the entangling surface to be a spatial $S^{d-2}$. It is known that the universal part of the entanglement entropy has the following property:
\eqn\ita{S_{EE} \propto a_d^*}
where $a_d^*$, defined in \refs{\MyersXS, \MyersTJ}, is a generalization of the $a$-type central charge to any dimension. In even dimensions, $a_d^*=a$, the coefficient of the $d$-dimensional Euler density in the conformal anomaly; in odd dimensions, $a_d^*\propto \log Z_{S^d}$, where $Z_{S^d}$ is the partition function of the CFT on the sphere. In $d=3$, for instance, $2\pi a_3^*=F$, the star of the recently discovered $F$-theorem \refs{\MyersTJ, \JafferisZI,\KlebanovGS}. The result \ita\ acts as a beautiful liaison between quantum entanglement and a fundamental quantity of any CFT; indeed, a proof of the $F$-theorem comes from the entanglement perspective \CasiniEI. 

On its face, the R\'enyi entropy away from the limiting value $q=1$ takes no such universal form. One is led to wonder whether this is completely true. That is, we seek new universal features of R\'enyi entropy that might allow us to more deeply understand the nature of entanglement or, conversely, foundations of conformal field theory.

We will show that the R\'enyi entropy has the following property near $q=1$:
\eqn\rya{ S'_{q=1} = -{\rm Vol}(\IH^{d-1})\cdot {\pi^{d/2+1}\Gamma(d/2)(d-1)\over(d+1)!}C_T }
where the prime denotes a derivative with respect to $q$. The constant $C_T$ is defined via the stress tensor vacuum two-point function in flat space \OsbornCR,
\eqn\ctm{\la T_{ab}(x)T_{cd}(0)\ra_{\IR^d} = C_T{I_{ab,cd}(x)\over x^{2d}}}
where $I_{ab,cd}(x)$ is a particular tensor structure which we recall below. To isolate the universal part of \rya, one simply substitutes the regulated volume of the hyperboloid.

Up to a choice of convention, \rya\ was effectively first conjectured in \HungNU\ and discussed again in \GalanteWTA. Those authors noticed patterns in the conformal dimensions $h_q$ of twist operators (calculated holographically) that can be used to compute R\'enyi entropy, and conjectured that $h'_{q=1}$ is universally controlled by $C_T$. These dimensions and the R\'enyi entropy stand in a fixed relation to each other, and hence so do $h'_{q=1}$ and $S'_{q=1}$.

The result \rya\ is based on the observation that for {\it any} entangling surface, the perturbative expansion of the R\'enyi entropy around $q=1$ is a sum of connected correlators of the entanglement Hamiltonian, defined by $\rho = e^{-H}/\Tr(e^{-H})$. Specializing to a spherical entangling surface, the proof of \rya\ relies on the work of \CasiniKV, where it was shown that one can conformally map the computation of R\'enyi entropy to the computation of the thermal partition function of a CFT living on $\IR\times \IH^{d-1}$, with inverse temperature $\b=1/T=2\pi q$ (here and henceforth we set the length scale to one). The R\'enyi entropy is just the thermal entropy of that CFT, and the volume factor in \rya\ is its spatial volume; furthermore, the entanglement Hamiltonian $H$ becomes, up to a unitary transformation, the thermal Hamiltonian of the hyperbolic CFT. The expansion around $q=1$ thus becomes a sum over connected correlators of the Hamiltonian on $S^1 \times \IH^{d-1}$ in the thermal state with $q=1$. At this special temperature, $S^1 \times \IH^{d-1}$ is conformal to $\IR^d$, and the thermal state maps to the vacuum of $\IR^d$. So these correlators can be calculated by conformally mapping the stress tensor back to $\IR^d$, and the result \rya\ follows from the two-point function. It bears noting that these thermal correlators are highly non-generic, given their relation to flat space correlators. 

Because of this connection between R\'enyi entropy and hyperbolic CFT partition functions, \rya\ is also a statement about general CFTs living on $\IR\times \IH^{d-1}$ at inverse temperature $\b=2\pi q$. Following \KlebanovUF, let us define the quantity
\eqn\cta{\cF_q = -\log Z_q~.}
Using standard thermodynamic relations relating $S_q$ and $\cF_q$ which we provide momentarily, \rya\ translates to the equation
\eqn\rytca{\cF''_{q=1} = -{\rm Vol}(\IH^{d-1})\cdot{2\pi^{d/2+1}\Gamma(d/2)(d-1)\over(d+1)! }C_T~.}
Again, the volume is meant to be regulated.

While our argument is purely field theoretic, it was inspired by holographic identifications of CFT R\'enyi entropy with hyperbolic black hole thermal entropy \refs{\CasiniKV, \HungNU, \BelinDVA}. We will review some calculations of \HungNU\ below, when we provide examples of \rya\ in action. It is also worth noting that in the limit $q\rar 0$, the R\'enyi entropy displays a different kind of universal behavior: it is controlled by the thermal free energy of the CFT living on the plane \SwingleHGA. 

We defer fuller reflection on \rya\ and \rytca\ to section 4 of the paper. In short, an obvious virtue of \rya\ is that we can apply knowledge of the quantity $C_T$ to the study of R\'enyi entropy (and hyperbolic CFT partition functions), and vice versa. Among the implications of this fact are that we can read off $S'_{q=1}$ in theories for which little about the R\'enyi entropy is known, including interacting CFTs that are not necessarily at large $N$; obtain $1/N$ corrections to $S'_{q=1}$ in large $N$ theories where results about $C_T$ are known; derive perturbative R\'enyi non-renormalization theorems; and impose and inform bounds obtained via the conformal bootstrap. Indeed, it seems to be quite challenging to compute R\'enyi entropy in a generic interacting CFT, so \rya\ is a welcome foothold. To exemplify some of these possibilities, we will make quantitative statements about R\'enyi entropy in the 3d $O(N)$ vector model and Chern-Simons deformations thereof \refs{\GiombiKC, \AharonyJZ}.

The holographic arguments of \CasiniKV\ further imply a dual gravitational version of the above statements. In particular, working solely from knowledge of $C_T$ in a CFT with a gravity dual, one can quantitatively determine quantum and stringy corrections to the thermal entropy of hyperbolic black hole solutions with linearized temperature fluctuations around $\b=2\pi$.

It is not yet clear to what extent the behavior of the R\'enyi entropy near $q=1$ can teach us about its behavior at arbitrary $q$, but we are hopeful that \rya\ will spur further insights.

The paper is organized as follows. In section 2 we establish \rya. Section 3 uses explicit calculations of R\'enyi entropy in the literature to provide a satisfying check of \rya\ in several examples, at weak and strong coupling and in even and odd dimensions. Section 4 presents various remarks on the consequences of this relation that we feel are especially exciting, and we close in section 5 by applying \rya\ to 3d vector models at large $N$, where we present new quantitative results for $S'_{q=1}$.

\newsec{A direct CFT argument}
 As stated in the introduction, we want to compute the R\'enyi entropy across $S^{d-2}$ of a $d$-dimensional CFT in its ground state. Our starting point is the fact that the R\'enyi entropy can be computed from the path integral of a CFT conformally mapped to $S^1\times \IH^{d-1}$ where the circle has length $\b=2\pi q$ \CasiniKV. We call this space $H^d_q$, and write its metric as
\eqn\ctj{\eqalign{ds^2_{H^d_q} &= d\t^2+du^2+\sinh^2 u ~d\Omega_{d-2}^2\cr}}
where $d\Omega^2_{d-2}$ is the line element on $S^{d-2}$, and the coordinates obey $u\in [0,\infty)$ and $\t\sim \t+2\pi q$. When $q=1$, we call the space $\cH^d\equiv H^d_{q=1}$, which is conformal to $\IR^d$,
\vskip .1 in
\eqn\ctja{\eqalign{ds^2_{\IR^d} &= dt^2+dr^2+r^2d\Omega_{d-2}^2\cr}}
where $r\in [0,\infty)$ and $t$ is Euclidean time. The explicit coordinate transformation relating $\cH^d$ to $\IR^d$ is
\eqn\ctja{t = {\sin \t\over \cosh u+\cos\t}~, ~~ r = {\sinh u\over \cosh u +\cos\t}}
whereby the two spaces are related as
\eqn\ctk{ds^2_{\IR^d} = \Omega^2ds^2_{\cH^d}~, \quad {\rm where} ~~ \Omega = {1\over \cosh u +\cos \t}~.}

The partition function on $H^d_q$ is
\eqn\ctb{Z_q = \Tr (e^{-2\pi q \hat{E}_q})}
and energy is defined as
\eqn\ctba{E_q = \int_{\IH^{d-1}} d^{d-1}x \sqrt{g}~\la T_{\t\t}(x)\ra_{H^d_q}~.}
Let us also write the energy as
\eqn\drc{E_q = {1\over 2\pi}\cF'_q= {\Tr(\hat{E}_qe^{-2\pi q \hat{E}_q})\over \Tr(e^{-2\pi q \hat{E}_q})}}
The relation between the reduced density matrix $\rho$ and the partition function $Z_q$ is
\eqn\dra{\Tr[\rho^q] = {Z_q\over (Z_1)^q}}
which leads to the following definition of the R\'enyi entropy in terms of $\cF_q$ \KlebanovUF:
\eqn\ctc{S_q = {q\cF_1-\cF_q\over 1-q}~.}
In the limit $q\rar 1$, one recovers the entanglement entropy, which is simply equal to the thermal entropy of the CFT. 

It is convenient to think in terms of the energy. If we expand \ctc\ around $q=1$ and use \drc, then defining $\dq\equiv q-1$, we get
\eqn\rytb{S_q = S_{EE} + 2\pi\sum_{n=1}^{\infty}{1\over (n+1)!} {\p^nE_q\over \p q^n}\Big|_{q=1}\dq^n~.}
This is just an expansion in energy correlators on $\cH^d$. So to compute the R\'enyi entropy in an expansion around the entanglement entropy value $q=1$, we must compute thermal stress tensor correlators in a CFT on $\IR \times \IH^{d-1}$ at $\b=2\pi$. The temperature fluctuation acts as a source for the stress tensor in the partition function, $Z[J]=\Tr(e^{-2\pi \hat{E}+J\hat{E}})$ with source $J=\delta \b=2\pi \dq$ and energy $\hat{E}\equiv \hat{E}_{q=1}$. Reasoning holographically, the temperature fluctuation appears as a metric perturbation of the hyperbolic black hole which couples to the boundary stress tensor. 

To prove \rya, we must consider the $n=1$ term in \rytb,
\eqn\cte{S'_{q=1}= -2\pi^2\left[{\Tr(\hat{E}\hat{E}e^{-2\pi \hat{E}})\over \Tr(e^{-2\pi \hat{E}})}- {\Tr(\hat{E}e^{-2\pi \hat{E}})\over \Tr(e^{-2\pi \hat{E}})} {\Tr(\hat{E}e^{-2\pi \hat{E}})\over \Tr(e^{-2\pi \hat{E}})}\right]}
which is a connected two-point function of the energy of the theory on $\cH^d$. Then \rya, written in path integral language, says that 
\eqn\dri{\eqalign{&\int_{\IH^{d-1}}d^{d-1}x\sqrt{g}\int_{\IH^{d-1}}d^{d-1}y\sqrt{g}~\la T_{\t\t}(x)T_{\t\t}(y)\ra_{\cH^d} -\left(\int_{\IH^{d-1}}d^{d-1}x\sqrt{g}~\la T_{\t\t}(x)\ra_{\cH^d}\right)^2\cr&= {\rm Vol}(\IH^{d-1}){\pi^{d/2-1}\over 2}{\Gamma(d/2)(d-1)\over (d+1)!}C_T~.}}
This state in which the expectation value is taken is not the ground state \refs{\EmparanPM, \EmparanGF}, but it has the special property that under a conformal transformation to $\IR^d$, it maps to the vacuum.

To summarize, our goal is to prove \dri. Our strategy is to conformally map the stress tensor from $\cH^d \mapsto \IR^d$. This forces us to address the conformal anomaly and the distinction between even and odd dimensions. 

For $d$ odd, there is no anomaly: the stress tensor is traceless for a CFT on a curved background, there is no vacuum energy, and the conformal map of the stress tensor from $\cH^d \mapsto \IR^d$ is non-anomalous. For $d$ even, the stress tensor of the CFT on $\cH^d$ remains traceless, but carries a nonzero energy density.  This implies that the stress tensor transforms anomalously under the conformal map, just as in $d=2$. Under a conformal transformation, we can write the stress tensor transformation as  
\eqn\ctl{T_{\a\b}(x) = \Omega^{d-2}{dX^{a}\over dx^{\a}}{dX^{b}\over dx^{\b}}T_{ab}(X) + {\cal S}_{\a\b}(x)}
The second term, ${\cal S}_{\a\b}(x)$, is the anomalous piece that is present in even $d$:
\eqn\ctla{\eqalign{d~{\rm odd}:&\quad {\cal S}_{\a\b}(x) = 0 \cr
d~{\rm even}:&\quad {\cal S}_{\a\b}(x) \neq 0 ~.\cr}}

For simplicity, we begin with $d$ odd.
\subsec{Odd dimensions}
As we just noted, in odd dimensions we have $\la T_{\t\t}(x)\ra_{\cH^d}=0$. 
We are free to put one of the operators in \dri\ at the origin; the integration gives us an overall ${\rm Vol}(\IH^{d-1})$, leaving us to show
\eqn\cti{\int_{\IH^{d-1}}d^{d-1}y\sqrt{g}\la T_{\t\t}(0)T_{\t\t}(y)\ra_{\cH^d} ={\pi^{d/2-1}\over 2}{\Gamma(d/2)(d-1)\over (d+1)!}C_T~.}

The precise definition of $C_T$ is via the stress tensor two-point correlator in the vacuum on $\IR^d$ \OsbornCR,
\eqn\ctm{\la T_{ab}(x)T_{cd}(0)\ra_{\IR^d} = C_T{I_{ab,cd}\over x^{2d}}}
where,
\eqn\ctn{I_{ab,cd}(x) = \half\Big(I_{ac}(x)I_{bd}(x)+I_{ad}(x)I_{bc}(x)\Big)-{1\over d}\delta_{ab}\delta_{cd}}
and
\eqn\cto{I_{ac}(x) = \delta_{ac}-2{x_cx_d\over x^2}~.}
The indices run over coordinates on $\IR^d$.

Now, we perform the coordinate transformation \ctja, transform the energy density $T_{\t\t}(x)$ using \ctl\ and \ctla, and subsequently evaluate \cti. Suppressing dependence on the sphere coordinates in what follows, we choose to put the second operator in \cti\ at time $\t=\pi$, whereby the operators are at opposite sides of the cylinder. This is a convenient choice: using the definitions \ctja\ and \ctk, one can show that $T_{\t\t}(0)$ and $T_{\t\t}(\t=\pi,u)$ transform simply as 
\eqn\ctp{\eqalign{T_{\t\t}(0) &=T_{tt}(0)\cdot  2^{-d}\cr
T_{\t\t}(\t=\pi,u) &= T_{tt}(t=0,r')\cdot (\cosh u-1)^{-d}}}
where
\eqn\ctq{r'={\sinh u\over \cosh u-1}~.}
Thus, using the formulae \ctm-\cto\ for the flat space correlator and the definition \ctq, one quickly computes
\eqn\ctr{\eqalign{\la T_{\t\t}(0)T_{\t\t}(\t=\pi,u)\ra_{\cH^d}&=  {1\over 2^d(\cosh u-1)^{d}}\la T_{tt}(0)T_{tt}(t=0,r') \ra_{\IR^d}\cr
&= {C_T(d-1)\over 2^dd} {(\cosh u-1)^{d}\over(\sinh u)^{2d}}~.\cr}}
We have written the correlator on $\IR^d$ in terms of coordinates on $\cH^d$ to ease the integration in \cti, which we now perform. The integration over $S^{d-2}$ gives the volume 
\eqn\ctt{{\rm Vol}(S^{d-2}) = {2\pi^{(d-1)/2}\over \Gamma\left({d-1\over 2}\right)}}
Putting this all together, one is left with
\eqn\ctu{\eqalign{\int_{\IH^{d-1}}d^{d-1}y\sqrt{g}\la T_{\t\t}(0)T_{\t\t}(y)\ra_{\cH^d} &= {2\pi^{(d-1)/2}\over \Gamma\left({d-1\over 2}\right)}{C_T(d-1)\over 2^dd}\int_{0}^{\infty}du~(\sinh u)^{-d-2}(\cosh u-1)^{d}\cr
&= {\pi^{(d-1)/2}\over 2^{d-1}\Gamma\left({d-1\over 2}\right)d(d+1)}C_T~.}}
The last line requires $d>1$, which is of course the case here.

Finally, to establish \cti\ we only need to show that
\eqn\ctv{{\pi^{(d-1)/2}\over 2^{d-1}\Gamma\left({d-1\over 2}\right)d(d+1)}C_T= {\pi^{d/2-1}\over 2}{\Gamma(d/2)(d-1)\over (d+1)!}C_T~.}
Cancelling a bunch of factors we are left with the equality
\eqn\ctw{\Gamma\left({d\over 2}\right)\Gamma\left({d-1\over 2}\right)=2^{2-d}\sqrt{\pi}~\Gamma(d-1)~.}
This is indeed a property of the Gamma function.

\subsec{Even dimensions}

In this case, $\la T_{\t\t}(x)\ra_{\cH^d}\neq 0$, and in fact the CFT on $H_q^d$ has nonzero energy at all finite $q$, analogous to the Casimir energy of a CFT on a spatial sphere, e.g. \refs{\BrownSJ, \CandelasGF}. What compensates for the nonzero one-point function in \dri\ is the anomalous piece of the stress tensor under the conformal mapping, ${\cal S}_{\a\b}(x)$. At this stage we need to know what this equals.\foot{We thank Per Kraus and Hugh Osborn for helpful discussions about this topic.} In $d=2$ this is essentially the Schwarzian derivative, but in higher dimensions its form is less familiar. Fortunately, we can argue as follows \HungNU. We are interested in the transformation of $T_{\t\t}(x)$ under the mapping \ctja, which we write as:
\eqn\ctl{T_{\t\t}(x)-{\cal S}_{\t\t}(x)= \Omega^{d-2}{dX^{a}\over d\t}{dX^{b}\over d\t}T_{ab}(X)~.}
In this case, $\lbrace x\rbrace$ parameterizes $\cH^d$, and $\lbrace X\rbrace$ parameterizes $\IR^d$. Taking an expectation value on both sides in the relevant states, we know that $\la T_{ab}(X)\ra_{\IR^d}=0$ by conformal invariance. So we have
\eqn\drf{\la {\cal S}_{\t\t}(x)\ra_{\cH^d}=\la T_{\t\t}(x)\ra_{\cH^d}~.}
That is, the one-point function of the anomalous term in \ctl\ is just the energy density of the CFT on $\cH^d$. 

As an operator statement, $S_{\t\t}(x)$ should equal the energy density times the identity operator:
\eqn\drg{{\cal S}_{\t\t}(x) = \la T_{\t\t}(x)\ra_{\cH^d}}
This generalizes to arbitrary tensorial indices. In fact, one can be explicit about the tensorial form of $S_{\a\b}(x)$ in conformally flat backgrounds: it is a covariantly conserved object whose trace reproduces the conformal anomaly. In particular, its trace is proportional to the $d$-dimensional Euler density times the $a$-type central charge (up to ambiguities involving total derivatives in the conformal anomaly that can be removed). We will not need them here, but there are explicit expressions for $S_{\a\b}(x)$ in the literature for $d=2,4,6$ \refs{\BrownSJ, \CappelliVW, \HerzogED}.\foot{We have assumed that $S_{\a\b}(x)$ is a pure curvature term, i.e. that one can always improve the stress tensor such that this is the case.}

Using \drg, the derivation is all but complete. Instead of \ctr, one has
\eqn\drh{\eqalign{\la T_{\t\t}(0)T_{\t\t}(\t=\pi,u)\ra_{\cH^d}&=  {1\over 2^d(\cosh u-1)^{d}}\la T_{tt}(0)T_{tt}(t=0,r') \ra_{\IR^d}\cr&+ \la T_{\t\t}(0)\ra_{\cH^d}\la T_{\t\t}(\t=\pi,u)\ra_{\cH^d}\cr}}
where we have used that $\la T_{tt}\ra_{\IR^d}=0$. The second line cancels the one-point contributions in the first line of \dri, and the rest of the argument proceeds as in odd dimensions.

\newsec{Explicit examples}

It is satisfying to verify that \rya\ is true in explicit examples, so let us perform the perturbative expansion of R\'enyi entropies in various theories in which it has already been computed. These simple checks are culled from holographic and pure field theoretic calculations. The reader may also find it useful to have an expression for the regulated hyperbolic volume, so as to extract the universal part of the results:
\eqn\ryfa{{\rm Vol}(\IH^{d-1}) = {\pi^{d/2}\over \Gamma(d/2)}\times \left\{ \matrix{&(-)^{d/2-1}{2\pi^{-1}}\log (R/\eps)\quad\quad  &d~{\rm even} \cr
 &(-)^{(d-1)/2} \quad &d~{\rm odd} } \right. }
$\eps$ is a short-distance cutoff, and $R$ is the radius of curvature of $\IH^{d-1}$. 

As a preliminary check, we note that \rya\ is trivial in two dimensions: the entangling sphere shrinks to zero size. One should instead consider the subspace to be a single line segment of length $R$. The single-interval R\'enyi entropy in the ground state of a two-dimensional CFT is given by the universal result first derived in \HolzheyWE,
\eqn\plla{S_q = {c_L+c_R\over 12}\left(1+{1\over q}\right)\log{R\over\eps}}
where $(c_L,c_R)$ are the left and right central charges. It is straightforward to reconcile \plla\ with our result \rya.\foot{One first passes to complex coordinates $(z,\zb)$ and writes the chiral components of the stress tensor as $T(z) = 2\pi T_{zz}$ and $\overline{T}(\zb) = 2\pi T_{\zb\zb}$, then uses equations \ctm\ -- \cto\ and \ryfa\ in conjunction with the CFT vacuum two-point function $\la T(z)T(0) \ra_{\IR^2} = c_L/2 z^4$ and its right-moving (antiholomorphic) counterpart.} As is well-known (see \HungNU\ for instance), such universality does not hold for the R\'enyi entropy of a ball in $d>2$, in analogy to the multiple-interval case in $d=2$; in the present context, this is made clear via the perturbative expansion around $q=1$ as a sum over connected $n$-point correlation functions of the stress tensor.

\subsec{$(d+1)$-dimensional Gauss-Bonnet gravity}
We refer to Section 2 of \HungNU, where the R\'enyi entropy $S_q$ in a $d$-dimensional CFT dual to Gauss-Bonnet gravity was computed holographically using hyperbolic black hole thermodynamics. We follow all conventions of \HungNU\ except for one which we spell out shortly. 

The action for this theory is \LovelockYV\
\eqn\ryb{S = {1\over 2\ell_p^{d-1}}\int d^{d+1}x\sqrt{-g}\left[R+{d(d-1)\over L^2}+{\l L^2\over (d-2)(d-3)}E_4\right]}
where 
\eqn\ryba{E_4=R_{\mu\nu\rho\sigma}R^{\mu\nu\rho\sigma}-4R_{\mu\nu}R^{\mu\nu}+R^2}
is the 4d Euler density, and $\l$ is the dimensionless higher derivative coupling. The Greek indices run over all bulk coordinates. 

The asymptotically AdS hyperbolic black holes are characterized by a length scale at infinity, $\tilde{L}^2 = L^2/\fin$, which is fixed by the equations of motion as
\eqn\ryc{1-\fin-\l\fin^2=0~.}
The physical root is that for which $\fin=1$ at $\l=0$, namely
\eqn\ryca{\fin = {1-\sqrt{1-4\l}\over 2\l}~.}
The black holes have inverse temperature $\b=2\pi q$. \HungNU\ re-parameterize the temperature in terms of a variable $\xq=r_h/\tilde{L}$, where $r_h$ is the position of the event horizon for the metric written in coordinates such that the AdS boundary sits at $r\rar\infty$. Relating these two definitions of temperature yields an equation that determines $\xq$:
\eqn\ryd{0={d\over \fin}\xq^4-{2\over q}\xq^3-(d-2)\xq^2+{4\l\fin\over q}\xq+(d-4)\l\fin~.}
When $q=1$ the spacetime is isomorphic to AdS, and $x_{1}=1$. This is the universal hyperbolic black hole solution that exists in any theory admitting an AdS solution.  

With these definitions in hand, we can present the R\'enyi entropy $S_q$ \HungNU:
\eqn\ryf{\eqalign{S_q = {\pi q\over q-1}{\rm Vol}(\IH^{d-1})\left({\tilde{L}\over \ell_p}\right)^{d-1}&\Bigg[{1\over \fin}(1-\xq^d)-{3\over d-3}(1-\xq^{d-2})-{d-1\over d-3}\l\fin(1-\xq^{d-4})\cr&+{d\over d-3}(1-4\l)\left({1\over 1-2\l\fin}-{\xq^d\over\xq^2-2\l\fin}\right)\Bigg]~.}}
${\rm Vol}(\IH^{d-1})$ is the area of the hyperbolic black hole horizon.

Now we introduce the dictionary relating the parameters $(\l,\fin)$ to central charges of a dual CFT \BuchelSK. All we will need is the definition of $C_T$,\foot{This is called $\widetilde{C}_T$ by \HungNU. In addition, in their section on Gauss-Bonnet gravity the authors of \HungNU\ use a convention such that $\widetilde{C}_T|_{d=4}=c$, where $c$ is defined via the conformal anomaly,
\eqn\rymb{\la T^a_a\ra = {c\over 16\pi^2}C_{abcd}C^{abcd}-{a\over 16\pi^2}E_4~.}
In the current paper we employ the convention $C_T^{\rm here} =  {(d+1)!\over (d-1)\pi^d}\widetilde{C}_T^{\rm there}$, which is widely used and is the original convention of \BuchelSK\ and \MyersJV.} which in our convention is
\eqn\rye{C_T = {(d+1)!\over \pi^{d/2}\Gamma(d/2)(d-1)}\left({\tilde{L}\over \ell_p}\right)^{d-1}[1-2\l\fin]~.}
Using \ryca, one can write $\l$ as follows:
\eqn\ryea{\l={1\over 4}-\left(\left({\ell_p\over\tilde{L} }\right)^{d-1}{\pi^{d/2}\Gamma(d/2)(d-1)\over2 (d+1)!}C_T\right)^2~.}

Using \ryca\ and \ryea, we want to solve \ryd\ for $x_q$ in series around $q=1$, then plug into \ryf\ and take a derivative with respect to $q$. Due to the factor of $q-1$ in the denominator of the definition of $S_q$, we must solve \ryd\ through second order in fluctuations. Writing 
\eqn\ryg{\xq = 1+\sum_{n=1}^{\infty}\xq^{(n)}\delta q^n}
we find
\eqn\ryh{\eqalign{&\xq^{(1)} = -{1\over d-1}\cr
&x^{(2)}_q = {d\over 2(d-1)^4}\left(2d^2-7d+5+\left({\tilde{L}\over \ell_p}\right)^{d-1}{2(d+1)!\over \pi^{d/2}\Gamma(d/2)}C_T^{-1}\right)~.}}
The expression for $x^{(2)}_q$ is much more complicated when expressed in terms of $(\l,\fin)$. 

Finally, expanding the R\'enyi entropy \ryf\ around $q=1$ and plugging in \ryca, \ryea\ and \ryh, the first derivative of the R\'enyi entropy evaluated at $q=1$ is simply 
\eqn\ryi{S'_{q=1} = -{\rm Vol}(\IH^{d-1})\cdot {\pi^{d/2+1}\Gamma(d/2)(d-1)\over (d+1)!}~C_T}
in agreement with \rya. One can check that the second derivative at $q=1$ shows no such simplicity when expressed in terms of CFT quantities $(a_d^*, C_T)$.

\subsec{$d=4$ quasi-topological gravity}
This is a five-dimensional bulk theory that augments the Gauss-Bonnet theory by a certain six-derivative term \MyersJV. It can be generalized to any $d\geq 4$ but we stick to $d=4$ here. Compared to Gauss-Bonnet gravity, there is one more parameter in the action, and correspondingly one more CFT parameter in the game, namely the coefficient of the third possible tensor structure in the stress-tensor three-point function, denoted $t_4$ \OsbornCR. We can write all bulk quantities in terms of the central charges $(a,c)$ and $t_4$. We note that
\eqn\qtg{C_T = {40\over \pi^4}c}
in the dual CFT.

Our procedure is identical to the previous case, now with slightly more complicated equations. Accordingly our presentation is streamlined, and we refer the reader to \HungNU\ where the hyperbolic black hole solutions and full R\'enyi entropy are provided.

The asymptotic length scale of the solutions is again $\tilde{L}^2=L^2/\fin$, where now
\eqn\qtb{1-\fin+\l\fin^2+\mu\fin^3=0~.}
The black hole temperature is again a function of $x_q$, which now satisfies the equation
\eqn\qta{0={2\over \fin}\xq^6-{1\over q}\xq^5-\xq^4+{2\l\fin\over q}\xq^3+{3\mu\fin^2\over q}\xq-\mu\fin^2~.}
Let us trade $(\mu,\l,\fin)$ for $(a,c,t_4)$:\foot{There are small typos in equations (2.58) and (2.59) of \HungNU, in the definitions of $\l\fin$ and $\fin$. One can check that without these corrections, \qtb\ is not satisfied.}
\eqn\qtc{\eqalign{\l\fin &= \half{{c\over a}(1+6t_4)-1\over 3{c\over a}(1+3t_4)-1}\cr
\mu\fin^2 &= {{c\over a}t_4\over 3{c\over a}(1+3t_4)-1}\cr
\fin &= 2{3{c\over a}(1+3t_4)-1\over 5{c\over a}(1+2t_4)-1}}}
Then solving \qta\ perturbatively using the expansion \ryg, we find
\eqn\ryha{\eqalign{&\xq^{(1)} = -{1\over 3}~, ~~ \xq^{(2)} = {2\over 27}\left(6+12t_4-{a\over c}\right)~.}}
One can check that at $t_4=0$, \ryha\ is consistent with the Gauss-Bonnet result \ryh\ at $d=4$.

Plugging this all into a perturbative expansion of $S_q$ (see equation (2.56) of \HungNU), and using \qtg, we find
\eqn\rym{S'_{q=1} =-{\rm Vol}(\IH^{3})\cdot {\pi^3\over 40} C_T}
in agreement with \rya.

Both \ryi\ and \rym\ were noticed in \HungNU, as statements about twist field dimensions (up to conventions).

\subsec{Free bosons and fermions}

We will need the relevant values of $C_T$: following the conventions of \OsbornCR,
\eqn\ryqa{C_T({\rm complex} ~\phi) = {2d\over d-1}{1\over S_d^2}}
where $S_d = 2\pi^{d/2}/\Gamma(d/2)$ is the volume of a unit $(d-1)$-sphere. For Dirac fermions,
\eqn\ryqb{C_T({\rm Dirac}~\Psi)= \left\{ \matrix{&{d\over 2S_d^2}2^{d/2}\quad &d={\rm even} \cr &{d\over 2S_d^2}2^{(d-1)/2} &d={\rm odd} } \right. }
This is consistent with our convention relating $C_T$ and $c$ in even dimensions: e.g. when $d=4$, we have
\eqn\ryqc{C_T({\rm complex} ~\phi) = {2\over 3\pi^4}={40\over \pi^4}\cdot {1\over 60}}
where we have used $c=1/60$ for a complex scalar field.

It is trivial to confirm \rya\ in even dimensions, for both bosons and fermions, where results are available, e.g. \refs{\CasiniKT, \DowkerRP, \DowkerBU}. In odd dimensions, fewer analytic calculations have been done; we focus on $d=3$, where the R\'enyi entropies for free bosons and fermions were computed by QFT path integral methods \KlebanovUF. The values of $C_T$ are in fact equal only in $d=3$, so \rya\ tells us that we should get
\eqn\rywa{S'_{q=1}= -{\rm Vol}(\IH^2)\cdot {\pi\over 128}}
for both complex bosons and Dirac fermions. 

In \KlebanovUF, $\cF_q$ was found to be
\eqn\ryv{\cF_q = {-{\rm Vol}(\IH^2)\over 2\pi}\times \left[-2\a\int_0^{\infty}dx ~x \tanh^{\a}(\pi x)\log(1-\a e^{-2\pi q x})+q{(7-\a)\zeta(3)\over 8\pi^2}\right]}
where $\a=1$ for a complex boson, and $\a=-1$ for a Dirac fermion. Recalling the relations \rya\ and \rytca\ which imply $\cF''_{q=1} = 2S'_{q=1}$, one confirms \rywa\ upon taking two derivatives. At all higher orders, the result depends on $\a$. For $N$ fields, simply multiply by $N$.

\subsec{$\N=4$ super-Yang-Mills at weak and strong coupling}

The R\'enyi entropies across $S^2$ are known in the weak $(\l=0)$ and strong $(\l=\infty)$ coupling limits of $\N=4$ super-Yang-Mills (SYM) \refs{\FursaevMP, \GalanteWTA}. The former are computed in perturbation theory, and latter are computed holographically from 5d Einstein gravity. In addition to checking the result \rya, we point out that non-renormalization of $c$ as a function of $\l$ \AnselmiAM\ implies the following two statements:\vs

1. $S'_{q=1}$ should be the same at weak and strong coupling.

2. Upon incorporating $\a'$ corrections to the Einstein gravity result, $S'_{q=1}$ should not change. \vs

The results for the universal part of $S_q$ -- before including stringy corrections -- are 
\eqn\ryn{\eqalign{S_q = -N^2s(q,\l)\log{R\over \eps}}}
where 
\eqn\ryoa{\eqalign{s_{\rm weak}(q)\equiv \lim_{\l\rar 0}s(q,\l) &= {1+q+7q^2+15q^3\over 24q^3} \cr
s_{\rm strong}(q)\equiv \lim_{\l\rar \infty}s(q,\l) &= {1+q\over 64q^3}{(5\sqrt{1+8q^2} -3)(1+\sqrt{1+8q^2})^2\over 3+\sqrt{1+8q^2}} \cr}}
The full function $s(q,\l)$ is not known. This result already includes the regulated volume of $\IH^3$, cf. \ryfa. In $\N=4$ SYM at large $N$, $a=c=N^2/4$, and $C_T = 40c/\pi^4$ in our convention. So we can write \ryn\ as
\eqn\ryn{\eqalign{S_q = s(q,\l)\left[{\rm Vol}(\IH^3)\cdot {\pi^3\over 20}C_T\right]~.}}
Taking a derivative, 
\eqn\ryo{\eqalign{s'_{\rm weak}(q=1)  =s'_{\rm strong}(q=1) =-\half~.}}
As expected the coefficient of $S'_{q=1}$ matches \rya\ and the non-renormalization is visible.\foot{Since $S_q$ can be expanded around $q=1$ in terms of derivatives of the energy of the CFT on $S^1\times \IH^{3}$, this non-renormalization should also be visible in the energy itself. The energies at weak and strong coupling were computed in \EmparanGF, see section 4.1 therein. One happily confirms that while the energies exhibit a different functional form in terms of $\b=2\pi q$, their first derivatives evaluated at $q=1$ are equal.}

In \GalanteWTA, the subleading corrections to the strong coupling result were incorporated via the $R^4$ correction to type IIB supergravity \GreenTV. The corrections to the hyperbolic black hole metrics induce a correction to $S_q$. All we need here is the $q$-dependence of this correction -- call it $s_{\rm strong}^{R^4}(q)$ -- which is
\eqn\ryp{s_{\rm strong}^{R^4}(q) \propto {(1+q)^4(1-q)^3\over q^3(3+\sqrt{1+8q^2})^4}~.}
This is clearly consistent with the non-renormalization of $C_T$. 

Before moving on, we must point out an open issue. We have established that the $n^{\rm th}$ derivative of $S_q$ evaluated at $q=1$ is fixed by the $(n+1)$-point connected correlator of the Hamiltonian on the hyperbolic cylinder $\cH^d$. One might then expect $S''_{q=1}$ to be unrenormalized in the case of $\N=4$ SYM -- indeed, any $\N=1$ superconformal field theory (SCFT) in four dimensions -- on account of the fact that the stress tensor three-point function on $\IR^4$ is also not renormalized as a function of $\l$ \refs{\ErdmengerYC, \AnselmiAM, \OsbornQU}. Using \ryoa\ one can check that this is not the case. There is no immediate problem, since \rytb\ involves correlators on $\cH^4$, so we expect this to be related to subtleties in the conformal mapping and the definition of the stress tensor. We do not resolve this issue here.

\newsec{Comments and consequences}

Let us offer up an assortment of qualitative remarks about the nature of our result, after which we make some new and quantitative statements about R\'enyi entropies in 3d vector models.

In even-dimensional CFTs in flat space, the universal part of the entanglement entropy across a general entangling surface is a known function of the geometry and embedding of the surface, and of the central charges appearing in the conformal anomaly \SolodukhinDH. When the surface is not a sphere, all central charges can appear. For a cylindrical surface, the entanglement entropy is proportional to linear combinations of $c$-type central charges. 
One novelty of \rya\ is that the {\it R\'enyi} entropy for a {\it fixed, spherical} entangling surface also contains definite information about all conformal anomaly coefficients. Our result, like \ita, also extends to odd dimensions. 

$C_T$ has received attention from the conformal bootstrap program, which  has experienced a resurgence of late \refs{\RattazziGJ,
\PolandWG,
\PolandEY,
\VichiUX,
\ElShowkHT,
\KosTGA}. Many CFTs appear to sit on the edges of allowed parameter space as determined by the bootstrap, in the sense that their spectra and central charges saturate bounds imposed by crossing symmetry. It is very intriguing to see connections between consequences of crossing symmetry and measurements of entanglement, and we would be eager to see them made explicit.  Can we understand the saturation of the bootstrap bounds from a quantum entanglement point of view, and what could the former imply about the allowed degree of correlations? 

In view of these thoughts, it is tempting to conjecture that the saturation of bootstrap bounds implies an extremization principle for R\'enyi entropy. According to our result \rya, this is evidently true for $S'_{q=1}$: minimization of $C_T$ implies maximization of $S'_{q=1}$. It is premature to say whether anything like this obtains more generally, but we hope that our result may stimulate further inquiry.

In 4d CFTs, $C_T\propto c$. There are bounds on the ratio $a/c$ in unitary 4d CFTs, and stricter bounds in SCFTs, coming from the demand of positivity of energy correlators \refs{\HofmanAR, \BuchelSK}.\foot{The definition of `energy correlator' used in those references is not exactly what we have computed in Section 2, but both refer to integrated stress tensor correlators.} In non-supersymmetric 4d CFTs, the bound is
\eqn\cma{1/3\leq a/c\leq 31/18}
and there are stronger bounds in supersymmetric theories. So the ratio $S_{EE}/S'_{q=1}$ is bounded both from above and below. 

In a general 4d, $\N\geq 1$ SCFT, the central charges $a$ and $c$ are not renormalized in the presence of exactly marginal deformations \AnselmiAM. Thus in these theories the $S'_{q=1}$ piece of the R\'enyi entropy is subject to a non-renormalization theorem, just as we explicitly showed earlier for $\N=4$ SYM. More generally, in any dimension, anytime $C_T$ is unrenormalized along a conformal manifold, we have a corresponding non-renormalization theorem for $S'_{q=1}$.

 Moving down one dimension, consider a 3d $\N=2$ SCFT: in such theories, $C_T\propto \t_{RR}$, the coefficient of the two-point function of the $R$-symmetry current. In \NishiokaHAA, a new observable called `super-R\'enyi entropy', $S_q^{\rm susy}$, was defined for these theories. A primary virtue of $S_q^{\rm susy}$ is that it can be computed exactly using localization, and while it shares no direct relation to ordinary R\'enyi entropy for generic $q$, it reduces to the entanglement entropy at $q=1$. $S_q^{\rm susy}$ has a specific relation to the partition function of the SCFT on a squashed $S^3$. Combining that fact with a result of \ClossetRU, the authors of \NishiokaHAA\ show that 
\eqn\nsa{\p_qS_{q=1}^{\rm susy}\propto \t_{RR}\propto C_T~.}
As $S_q^{\rm susy}\neq S_q$, our results are distinct, but complementary.

It took little work to show that the perturbative expansion of the R\'enyi entropy around $q=1$ is determined by stress tensor correlation functions. Beyond two points, there are multiple tensor structures that appear in these correlators. In sufficiently symmetric theories, there are further constraints on the linear combinations of structures that appear. To the extent that this is true for a given theory, the R\'enyi entropy thereof will be non-generic and perhaps more easily computable. In other words, the global symmetries of a CFT constrain its entanglement spectrum.

In a 3d CFT with higher spin symmetry broken by quantum effects, for example, the form of three-point functions of arbitrary higher spin currents is highly constrained: the coefficients of the three allowed tensor structures are known functions of the effective coupling and central charge of the theory \MaldacenaSF. It is reasonable to presume that this extends to all $n$-point functions, and perhaps to higher dimensions \refs{\DidenkoTV, \StanevQRA, \AlbaYDA}. Then the R\'enyi entropy in such theories is unusually constrained as well, at least in an expansion around $q=1$. Similar statements apply to energy correlators in 4d CFTs with values of $a/c$ that saturate the bounds \cma\ \ZhiboedovOPA. %

Before introducing any regulator for the UV divergences, the R\'enyi entropy is a monotonically decreasing function of $q$ \zyc:
\eqn\cca{S'_q\leq 0~.}
This is true for all $q$. In a holographic context, this result was explained by the positivity of the specific heat of the hyperbolic black hole \HungNU, which is also visible from \rytb. In the neighborhood of $q=1$, the positivity of $C_T$ in a unitary CFT provides yet another explanation. 

We offer one final comment before moving onto 3d vector models. It is interesting to run arguments from boundary to bulk and ask what we can learn about hyperbolic black hole entropy, as suggested in the introduction. If we consider the case where 4d hyperbolic black holes are dual to thermal states of 3d $\N=2$ SCFTs, we can, in the SCFT, compute $F$ by localization. This can be interpreted in the bulk as the exact quantum black hole entropy for the hyperbolic black hole at the special temperature $\b=2\pi$. This is a striking consequence of localization in the context of AdS/CFT. While this gives us much more information than we can currently process in the bulk, one would hope that we could use this wealth of information about 3d SCFTs emerging from localization techniques to learn something about quantum gravity.

\newsec{R\'enyi entropies of 3d vector models and Vasiliev theory}

There is a simple bulk argument for why the 3d free singlet boson and fermion theories have equal values for $S'_{q=1}$. These  theories are dual to the 4d Vasiliev theory of higher spin fields in AdS$_4$ \refs{\VasilievBA, \KlebanovJA, \GiombiMS}, which contains a free phase in its equations of motion, labeled $\theta_0$. The bulk duals of these free CFTs differ only through the choice of boundary condition on the  scalar field and the value of $\theta_0$. This phase is invisible at the free field level, though it appears in holographic computations of three- and higher-point functions. Because the linearized temperature perturbations we are considering only involve quadratic metric fluctuations, one concludes that the phase $\theta_0$ and the scalar boundary condition, and hence the distinction between the free singlet boson and fermion theories, are invisible at this order in the calculation. This completes the argument.

Now, we can also consider the exactly marginal coupling of a Chern-Simons gauge field to either the free complex boson or fermion CFTs, so that we now have $N$ complex bosons or fermions coupled to a $U(N)_k$ gauge field \refs{\GiombiKC, \AharonyJZ}. (One can have an $O(N)_k$ gauge field instead, corresponding to real scalars and fermions.) This parity-breaking deformation classically preserves the higher spin symmetry of the free theories in the 't Hooft limit in which $(N,k)\rar\infty$ with $\l=N/k$ fixed. Consequently, the family of Chern-Simons-fundamental matter theories in the 't Hooft limit is conjecturally dual to a continuous family of Vasiliev theories parameterized by the value of $\theta_0$. The holographic dictionary tells us that $\theta_0$ is linear in the CFT 't Hooft coupling $\l$ \ChangKT.\foot{We must mention that there is an entire function's worth of freedom, unconstrained by classical higher spin symmetry, in specifying the interactions among the Vasiliev master fields \VasilievBA. The way this `interaction ambiguity' fits into the vector model dualities described here is poorly understood; in particular, the field theories seem to suggest that this function can be reduced to a pure phase. } 

While $\theta_0$ only affects three-point functions and beyond, the Chern-Simons deformation is also visible in the central charge of the CFT, namely in the value of $C_T$. The results of \refs{\MaldacenaSF} imply that $C_T$ is a nontrivial function of $\l$, and this function was computed explicitly in \AharonyNH\ for the $U(N)_k$ Chern-Simons-boson theory:%
\eqn\rywb{C_T = {N\sin\pi\l\over \pi\l}\cdot C_T({\rm complex}~\phi)~.}
We provided the value of $C_T({\rm complex}~\phi)$ in \ryqa. The statement in the bulk is that in addition to the shift of $\theta_0$ induced by nonzero $\l$, the effective bulk Newton's constant of the Vasiliev theory is a function of $\l$ (and $N$).

The R\'enyi entropies for these theories have not yet been computed and it would be interesting to do so. While the full R\'enyi entropy may be complicated, we now know that for the $U(N)_k$ Chern-Simons-boson theory,
\eqn\rywk{
S'_{q=1} = -{\rm Vol}(\IH^2)\cdot {N\sin\pi\l\over 128\l}~.}
Analogous statements apply to the dual fermionic theory related to this one by 3d bosonization \refs{\AharonyNH, \AharonyNS}. The entanglement entropy $S_{EE}(\l)=-F(\l)$ has also not yet been calculated, but should be an even function\foot{The theory is formally invariant under a parity transformation combined with a sign flip of $\l$. The free energy is parity-invariant, which implies that $S_q$ is an even function of $\l$.}  of $\l$ with a smooth $\l\rar0$ limit to the free result given in \KlebanovUF, at least once a pure Chern-Simons contribution (of order $N^2$) has been accounted for. We expect this calculation to be tractable, at least perturbatively in $\l$. A similar calculation was done in \KlebanovTD\ for a $U(1)_k$ gauge field coupled to a large $N$ boson sector. 

We can also make statements about $1/N$ corrections to R\'enyi entropy in large $N$ field theories. For instance, one can ask, `what is the change in the R\'enyi entropy induced by the double trace flow from the free to critical fixed points of the $O(N)$ vector model?' This question was answered by \KlebanovGS\ for the entanglement entropy at large $N$ but not for the R\'enyi entropy, but we can partially answer this question now.

Let us phrase the following in terms of $\cF_q$, where we recall that $\cF_1=F$. The flow between fixed points of the $O(N)$ vector model is triggered by a double trace deformation of the UV action,
\eqn\dba{\delta S^{UV} \sim \int d^3x \sqrt{g}~ J^{(0)}J^{(0)}}
where the scalar singlet operator $J^{(0)}=\phi^i\phi^i$ has $\Delta=1$ in the UV. The resulting change in $F$ is \KlebanovGS\foot{Here and below, we follow \KlebanovGS\ in plugging in the regulated volume ${\rm Vol}(\IH^2)=-2\pi$.}%
\eqn\dbb{\delta F\equiv F^{UV}-F^{IR} =-{\zeta(3)\over 8\pi^2}+O(1/N)~.}
This was recently reproduced from a bulk computation in the Vasiliev theory on AdS$_4$ \GiombiYVA.

The full calculation of $\delta \cF_q\equiv \cF^{UV}_{q}-\cF^{IR}_q$ would involve computing the change in the partition function on $H^3_q$ due to the deformation \dba. Such a calculation, expanded in $\dq$, should yield
\eqn\pra{\eqalign{\delta \cF_q&\approx -{\zeta(3)\over 8\pi^2}+\left({5\over 144}+O(1/N)\right)\delta q^2 + O(\delta q^3)~.\cr}}
We have used equation \rytca\ and the result from \PetkouAD\ (see also \PetkouVU) that for a real scalar,
\eqn\prb{C_T^{UV}-C_T^{IR}= {5\over 12\pi^4}+O(1/N)~.}
We also used the fact that the vacuum energies on $\cH^{3}$ vanish. 
Using \ctc, this amounts to a determination of the $O(\dq)$ change in R\'enyi entropy  between the fixed points; in the spirit of \SachdevPR\ it should be noted that the $O(\dq^2)$ term of \pra, to leading order in $N$, is rational. In writing \pra\ we have presumed that any potential phase transitions in the critical $O(N)$ model on $H^3_q$ do not affect this determination of $\delta \cF_q$ near $q=1$. 

Very recently, conformal bootstrap methods were applied to the critical $O(N)$ vector model \KosTGA. One lesson is that this theory, like many before it, seems to saturate certain lower bounds on $C_T$ imposed by crossing symmetry, for all $N$. The intersection between this result, the implications for the bulk Vasiliev theory, and now R\'enyi entropy appears to be a rich one.

\vskip .3in

\noindent
{ \bf Acknowledgments}

\vskip .3cm

We thank Per Kraus for collaboration in the early stages of this work. We also thank Clay Cordova, Gary Gibbons, Guy Gur-Ari, Igor Klebanov and Ran Yacoby for helpful comments, and especially Per Kraus, Hugh Osborn and Mukund Rangamani for helpful comments and a reading of an earlier draft. Finally, we thank the Centro de Ciencias de Benasque Pedro Pascual for hospitality during the course of this work. The author has received funding from the European Research Council under the European Union's Seventh Framework Programme (FP7/2007-2013), ERC Grant agreement STG 279943, “Strongly Coupled Systems”.

\listrefs
\end